\documentclass[12pt,aps,pre,amsmath,amssymb,onecolumn]{revtex4}
\usepackage{graphicx}
\usepackage{dcolumn}
\usepackage{bm}
\usepackage{xcolor}
\usepackage{times}
\begin{document}
\title{Kortweg de-Vries solitons on electrified liquid jets}
\author{Qiming Wang}
\affiliation{Department of Mathematics and Statistics,
York University, Ontario M3J 1P3, Canada}
\author{Demetrios T. Papageorgiou}
\affiliation{Department of Mathematics, Imperial College London, South Kensington Campus, London SW7 2AZ, U.K.}
\author{Jean-Marc Vanden-Broeck}
\affiliation{Department of Mathematics, University College London, Gower Street, London WC1E 6BT, U.K.}

\begin{abstract}

The propagation of axisymmetric waves on the surface of a liquid jet under the action of
a radial electric field is considered. The jet is assumed to be inviscid and perfectly conducting,
and a field is set up by placing the jet concentrically inside a perfectly cylindrical tube whose wall
is maintained at a constant potential. A nontrivial interaction arises between the hydrodynamics
and the electric field in the annulus, resulting in the formation of electrocapillary waves.
The main objective of the present study is to describe nonlinear aspects of such axisymmetric waves
in the weakly nonlinear regime which is valid for long waves relative to the undisturbed jet radius.
This is found to be possible if two conditions hold: the outer electrode radius is not too small, and the
applied electric field is sufficiently strong. Under these conditions long waves are shown to be dispersive
and a weakly nonlinear theory can be developed to describe the evolution of the disturbances.
The canonical system that arises is the Kortweg de-Vries equation with coefficients that
vary as the electric field and the electrode radius are varied. Interestingly, the coefficient of the highest
order third derivative term does not change sign and remains strictly positive, whereas the coefficient $\alpha$ of
the nonlinear term can change sign for certain values of the parameters. This finding implies that solitary
electrocapillary waves are possible; there are waves of elevation for $\alpha>0$ and of depression for $\alpha<0$.
Regions in parameter space are identified where such waves are found.

\end{abstract}

\maketitle

\section{Introduction}\label{introduction}

Cylindrical liquid jets that support surface tension are susceptible to long wave instabilities: any linear perturbation
with wavelength longer than the jet circumference is unstable and eventually leads to the breakup of the jet
into drops - see Plateau \cite{Plateau}, Rayleigh \cite{Rayleigh1878}. Axisymmetric perturbations are
found to be the most dangerous ones in the absence of rotation or other external effects.
The instability is present for both low and high viscosity fluids due to its physical origin - a perfectly cylindrical liquid
thread can minimise its surface energy by disintegrating into spherical droplets of equivalent volume.
The stabilisation of such Rayleigh or capillary instabilities has
been the subject of numerous studies that invoke different physical mechanisms to influence the spectrum.
Of particular note are mechanisms involving electric and magnetic fields. 

Electric DC fields acting along the axis of the jet, have been observed to stabilise the Rayleigh instability and enable the formation
and persistence of longer liquid bridges, for example - see the experiments and theory of Raco \cite{Raco}, Ramos et al. \cite{Ramos94},
Burcham and Saville \cite{Burcham}. These observations were confirmed by numerical solutions of the full equations
by Volkov et al. \cite{Volkov}. In the presence of radial electric fields (as in the present study), the jet or liquid cylinder can be
stabilized or destabilized depending on the strength of the field and the separation distance between the liquid
and the outer cylindrical electrode. To fix things we consider a perfectly conducting liquid jet inside a concentric perfectly cylindrical
electrode so that the field acts in the annulus alone. This problem has a long history starting with linear studies by, for example,
Basset \cite{Basset}, Schneider et al. \cite{Schneider}, Neukermans \cite{Neukermans} and Artana et al. \cite{Artana1997,Artana1998}
who show that the electric field increases the critical Weber number (analogously the effective jet velocity) 
below which absolute instability is supported -
this may be useful in experimental studies since non-electrified jets support absolute instabilities at such low speeds that the
experiments become very delicate - see Chauhan et al. \cite{CMPR2006}. The manifestation of the instability into the
nonlinear regime and eventual jet pinching has been considered by Setiawan and Heister \cite{SH1997} using
boundary-element time-dependent computations. In this paper we are also study nonlinear aspects of the problem
and in particular we consider nonlinear waves valid in a electric-field induced long-wave stability window. 

Our work is related
to the study of Grandison et al. \cite{GrVaPaMiSp2007} who computed travelling waves of arbitrary amplitude and
sufficiently short wavelengths so that they are linearly stable due to capillarity. They considered both perfectly conducting
liquid jets, but also perfect dielectric (insulating) ones. In the latter case, a rod electrode needs to be present
along the axis of the configuration surrounded by a liquid annulus which is in turn surrounded by a second annular region
adjacent to the outer cylindrical electrode (such flow geometries have been studied in different non-electrified setups to
evaluate the effect of surfactants on thread-annular flows - see for example Bassom et al. \cite{BBP2012} and references
therein). The computations in \cite{GrVaPaMiSp2007} are electrified extensions of nonlinear cylindrical travelling waves
calculated by Vanden-Broeck et al. \cite{VMS1998}.

The problem of waves on the surface of liquid jets made of ferrofluids has also received attention over the years.
A magnetic field in the azimuthal direction is generated by passing a current through a thin wire placed on the axis
of the ferrofluid. It has been shown that a sufficiently strong induced magnetic field can stabilize capillary instabilities
and ultimately produce nonlinear travelling waves. These are governed by a Kortweg de-Vries (KdV) equation for
weakly nonlinear axisymmetric deformations - see Bashtovoi et al. \cite{Bashtovoi}, Rannacher and Engel \cite{Rannacher}.
Recent experiments by Bourdin et al. \cite{Bourdin} have confirmed the existence of axisymmetric depression
and elevation solitons that follow KdV dynamics.
In a recent theoretical study by Blyth and Parau \cite{BlythParau} solitary waves of arbitrary amplitude were computed
numerically and the elevation and depression weakly nonlinear solitons found in \cite{Rannacher} were calculated,
along with new branches of solitary waves. Large amplitude waves can develop to form toroidal trapped bubbles
as seen by Grandison et al. \cite{GrVaPaMiSp2007} in a different physical and mathematical setup. The present study
(as well as that in \cite{GrVaPaMiSp2007}) is more complicated mathematically than those for ferrofluids. The reason
for this is that the ferrofluid equations simplify significantly in axisymmetric geometries and the mathematical
problem is modified by adding a term to the Bernoulli equation that is inversely proportional to the local jet radius.
In the present case the problem is also axisymmetric but electric fields act in the annular region and need to be solved together
with the hydrodynamic problem to determine the interfacial position.
Nonetheless, results that are analogous to those for ferrofluids emerge, namely
weakly nonlinear KdV type dynamics along with depression and elevation solitons being supported.

The rest of the paper is organized as follows. Section \ref{formulation} formulates the mathematical model
and nonlinear boundary conditions, and also presents the linear dispersion relation for arbitrary wave numbers.
Section \ref{sec:weakly} constructs a long wave weakly nonlinear theory for the coupled electrohydrodyanmic
problem that leads to a Kortweg de-Vries type equation that can support depression or elevation waves depending
on the relative values of the electric field strength and the electrode radius; a phase diagram is calculated that separates
depression from elevation waves in the electric field - electrode radius space. Section \ref{conclusions} contains
some concluding remarks and possible future work.

\section{Problem formulation}\label{formulation}

\begin{figure}
\includegraphics[width=10cm, height=7cm]{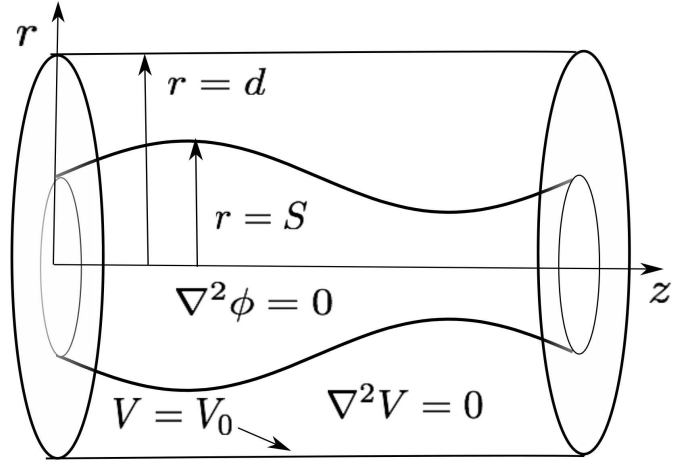}
\caption{Schematic of the problem showing the liquid thread placed concentrically inside a cylindrical electrode of radius $d$.}
\label{fig:schema}
\end{figure}

Consider an inviscid and incompressible liquid jet of density $\rho$ and undisturbed radius $a$.
The fluid is assumed to be a perfect conductor held at zero voltage and is concentrically placed inside a cylindrical
electrode of radius $d>a$. The outer electrode is maintained at a constant voltage potential $V_0$ so that
an electric field is set up in the dielectric annular region between the jet surface and the electrode - the annular
region is assumed to be hydrodynamically passive and can be taken to be air, for example,
with electric permittivity $\epsilon_0$. Considering axisymmetric deformations
and utilizing cylindrical polar coordinates $(r, \theta, z)$, we denote the evolving jet surface by $r=S(z,t)$.
The hydrodynamic problem in the region $0<r<S(z,t)$ is coupled with the electrostatic one in $S(z,t)<r<d$, through the
normal stress balance that is modified by the electrical Maxwell stresses at the interface; the effect consequently appears 
in the Bernoulli equation as stated below - for details of such derivations see \cite{WMPapa09,WangPapa2011JFM,GrVaPaMiSp2007},
for example. 
A schematic of the problem is given in Figure \ref{fig:schema}.

The flow is irrotational and so the velocity field is given by $\mathbf{u}=\nabla\phi$ where $\phi(r,z,t)$ is the fluid potential.
Incompressibility then implies that $\phi$ is harmonic. The electric field is given by $\mathbf{E}=-\nabla V$ where $V(r,z,t)$
is the electrostatic potential,
hence the field equations are
\begin{eqnarray}
\nabla^2 \phi=0,\qquad 0<r<S(z,t),\label{eq:LapPhi}\\
\nabla^2 V=0,\qquad S(z,t)<r<d,\label{eq:LapV}
\end{eqnarray} 
the latter equation arising from Gauss's law ${\rm{div}}(\epsilon_0\mathbf{E})=0$.
At the jet axis $r=0$, we impose regularity of $\phi$ while at the interface $r=S(z,t)$ we need to satisfy a kinematic condition
as well as the Bernoulli equation, namely
\begin{equation}
\phi_r=S_t+\phi_z S_z,\label{eq:BC1}
\end{equation}
\begin{eqnarray}
\rho \phi_t+\frac{1}{2} \rho \big(\phi_r^2+\phi_z^2 \big)
-\frac{(\epsilon_0/2)}{ (1+S_z^2) } \big(V_r -S_z V_z \big)^2=
-\frac{\gamma}{ (1+S_z^2)^{3/2} } \left[ \frac{1+S_z^2}{S}-S_{zz}\right]+K,\label{eq:Bernoulli}
\end{eqnarray}
where $K$ is a constant and $\gamma$ is the surface tension coefficient. Finally, the boundary conditions for the voltage potential $V$ are
\begin{eqnarray}
V=0,\qquad&{\rm on}&\qquad r=S(z,t),\label{bc:V1}\\
V=V_0,\qquad&{\rm on}&\qquad r=d.\label{bc:V2}
\end{eqnarray}
This completes the mathematical statement of the problem; we note that the hydrodynamic and electrostatic fields \eqref{eq:LapPhi}
and \eqref{eq:LapV} are coupled through the Maxwell stresses appearing in the Bernoulli equation boundary condition \eqref{eq:Bernoulli}.

\begin{figure}
\includegraphics[scale=0.7]{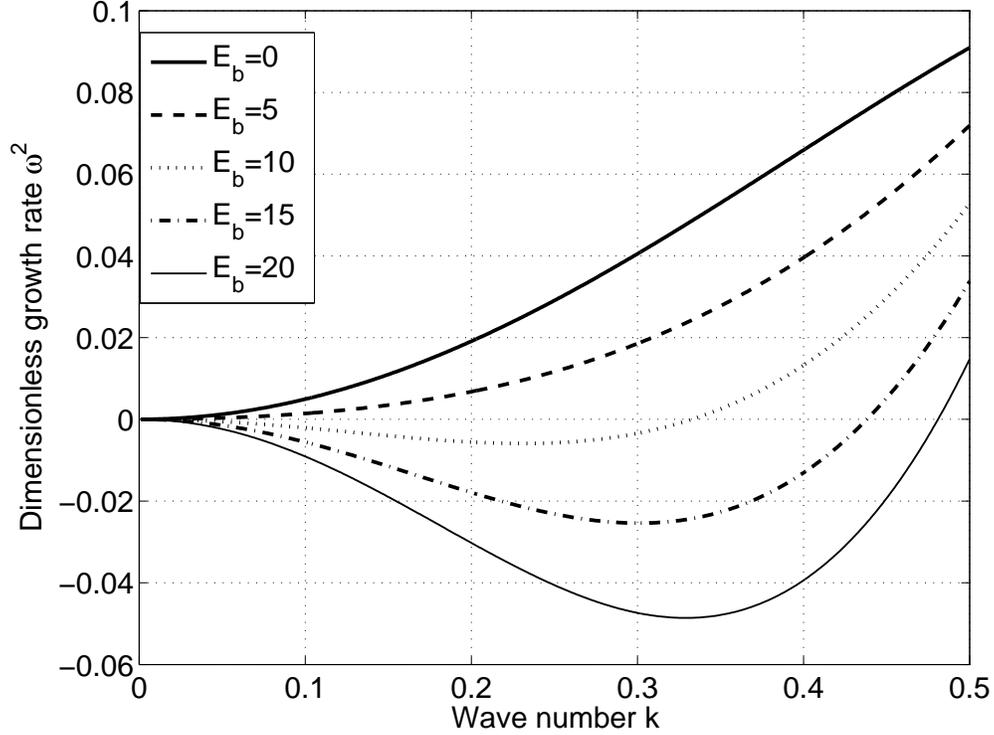}
\caption{Linear dispersion relation \eqref{invdisp0} showing stabilization of long waves due to the presence of
an electric field; the outer electrode is at $d=5a$. In the absence of a field, $E_b=0$, the classical Rayleigh instability is found - thick solid curve, while
the waves become dispersively stable as $E_b$ increases from $5$ to $20$ as shown.}
\label{fig:1}
\end{figure}

The following is an exact solution of the system \eqref{eq:LapPhi}-\eqref{bc:V2}:
\begin{eqnarray}
\mathbf{u}=\mathbf{0},\qquad S(z,t)=a,\qquad V=V_0\,\frac{\ln(r/a)}{\ln(d/a)}.\label{eq:exact}
\end{eqnarray}
This solution corresponds to a quiescent jet (note that any constant axial flow can be removed by a Galilean
transformation) of uniform radius $a$, and with a perfectly radial electric field in the annular region $a<r<d$. In the absence of an electric field
the perfectly cylindrical interface is susceptible to the Rayleigh-Plateau instability \cite{Rayleigh1878} - all linear disturbances with wavelengths longer
than the unperturbed jet circumference are unstable, and sufficiently short waves are stable. Note that analogous
results hold for viscous jets also, the main difference being in the magnitude of growth rates and their dependence on
additional parameters - see the pioneering work of Rayleigh \cite{Rayleigh1892}. In the presence of a radial electric field, Huebner \& Chu
\cite{HuChu} have derived the modified dispersion relation for disturbances proportional to $\exp(ikz+\omega t)$ 
with $k$ the wave number and $\omega$ the growth rate, which in our notation reads (in dimensional variables)
\begin{equation}\label{invdisp0}
 \left(\frac{\rho a^3}{\gamma}\right)\,\omega^2=\frac{ka\,I_1(ka)}{I_0(ka)}\left[1-(ka)^2-\frac{E_b}{\ln^2(d/a)}\left(1+ka\frac{K_{0}(kd)I_{1}(ka)+
 I_{0}(kd)K_{1}(ka)}{I_{0}(ka)K_{0}(kd)-I_{0}(kd)K_{0}(ka)}\right)\right],
\end{equation}
where 
\begin{eqnarray}
E_b=\frac{\epsilon_0 V_0^2}{\gamma a},\label{eq:Eb}
\end{eqnarray}
is a dimensionless parameter measuring the strength of the electric field relative to
capillary forces - it can be thought of as an electric Bond number.
When $E_b=0$, the classical Rayleigh instability result follows; the presence of a field can stabilize long waves and this can be seen by
considering \eqref{invdisp0} for small wave numbers $ka\ll 1$. The asymptotic result is
\begin{equation}
\left(\frac{\rho a^3}{\gamma}\right)\,\omega^2=\frac{k^2a^2}{2}\left(1-\frac{E_b[\ln(d/a)-1]}{\ln^3(d/a)}\right)+\ldots,\label{eq:disp}
\end{equation}
and we observe that whenever $({d}/{a})>{\rm e} \simeq 2.7183$ and $E_b$ is
sufficiently large, then
$\omega^2<0$, i.e. linear waves are stable and dispersive. 
Physically this requires the outer electrode to be sufficiently far from the undisturbed liquid surface and the applied
electric field to be sufficiently strong - these findings are similar to problems where  
the liquid jet is highly viscous \cite{WMPapa09}, or even with an annular fluid included \cite{WangPapa2011JFM}. 
For completeness in Figure \ref{fig:1} we provide plots of the dispersion relation \eqref{invdisp0} as $E_b$ varies for a fixed value of $d/a=5$;
the results clearly show the existence of long wave instability in the absence of an electric field, and the emergence and enhancement
of long wave dispersive stabilization as $E_b$ is increased.
Our aim in the remainder of this study is to describe the nonlinear dynamics
in the presence of such dispersive effects and to derive Kortweg-de Vries type
equations that describe electrocapillary waves in cylindrical liquid threads.

\section{Weakly nonlinear theory and derivation of Kortweg de-Vries equations}\label{sec:weakly}

In this section we construct nonlinear long wave solutions to the system \eqref{eq:LapPhi}-\eqref{bc:V2}.
In particular we assume that the typical axial wavelength of interfacial deformations, $\ell$ say, is long compared
to the undisturbed jet radius $a$, so that the slenderness ratio
\begin{equation}
\epsilon=\frac{a}{\ell}\ll 1.\label{eq:epsilon}
\end{equation}
We non-dimensionalize the problem using the following scalings
\begin{equation}
\phi=\left( \frac{\gamma \ell^2}{\rho a} \right)^{1/2} \,\phi^\prime, \quad V=V_0 \,V^\prime, \quad r=a\,r^\prime, \quad z=\ell\,z^\prime, \quad
S=a\,S^\prime,\quad t=\left( \frac{\rho a \ell^2}{\gamma}\right)^{1/2}\, t^\prime.\label{eq:scalings}
\end{equation}
Substituting \eqref{eq:scalings} into \eqref{eq:LapPhi}-\eqref{bc:V2} and dropping the primes, leads to the following
equations and boundary conditions - the small parameter $\epsilon$ enters the problem and will be utilized in the
development of asymptotic solutions later. (Note also that outer electrode radius $d$ becomes $D=d/a$ when non-dimensionalized.)

\begin{equation}
\phi_{rr}+\frac{1}{r}\phi_{r}+\epsilon^2\phi_{zz}=0,\qquad r=S(z,t),
\label{basic1}
\end{equation}

\begin{equation}
V_{rr}+\frac{1}{r} V_r+\epsilon^2 V_{zz}=0,\qquad S(z,t)<r<D,
\label{basic2}
\end{equation}
The boundary conditions for the fluid potential at the interface $r=S$ are the kinematic and dynamic conditions
\begin{equation}
\phi_r=\epsilon^2(S_t + \phi_z S_z)
\quad \hbox{on} \quad r=S
\label{eq:kin}
\end{equation}

$$
 \phi_t+ \frac{1}{ (1+\epsilon^2 S_z^2)^{3/2} } \left[ \frac{1+\epsilon^2 S_z^2}{S}-\epsilon^2 S_{zz}\right]-\frac{1}{2} 
E_b
\frac{1}{ (1+\epsilon^2 S_z^2) } \left(  \frac{\partial V} {\partial r} -\epsilon^2 S_z \frac{\partial V}{\partial z} \right)^2+
$$
\begin{equation}
\frac{1}{2}  \left[ \frac{1}{\epsilon^2}\left(\frac{\partial \phi}{\partial r}\right)^2+\left(\frac{\partial \phi}{\partial z}\right)^2 \right]=K
\qquad \hbox{on} \qquad r=S(z,t),
\label{basic5}
\end{equation}
where $K$ is dimensionless analogue of the constant appearing in \eqref{eq:Bernoulli}, and the electric Bond number parameter $E_b$
has been defined in \eqref{eq:Eb}.

The boundary conditions for the voltage $V$ are
\begin{equation}
V=0 \qquad \hbox{on} \qquad r=S,
\label{basic4b}
\end{equation}
\begin{equation}
V=1\qquad \hbox{on} \qquad r=D,
\label{basic4}
\end{equation}
where the former states that the jet interface is an equipotential since the fluid is a perfect conductor, while
the second condition corresponds to the prescribed voltage at the outer wall. 

The dimensionless system \eqref{basic1}-\eqref{basic4} is exact but contains a small parameter $\epsilon$.
Next we seek solutions for small $\epsilon$ with weakly nonlinear interfacial deformations, i.e. we write
\begin{equation}
S(z,t)=1+\epsilon^2 \eta(z,t)
\label{Sexp}
\end{equation}
and assume the expansions
\begin{equation}
V(r,z,t)=V_0(r,z,t)+\epsilon^2 \,V_1(r,z,t)+\epsilon^4\, V_2(r,z,t)+\dots
\label{Vexp}
\end{equation}
\begin{equation}
\phi(r,z,t)=\epsilon^2\, \phi_0(r,z,t)+\epsilon^4\, \phi_1(r,z,t)+\dots.
\label{phiexp}
\end{equation}
We also introduce the canonical Korteweg de-Vries  scaling
\begin{equation}
x=z-ct, \qquad \tau=\epsilon^2  t,
\label{scaling}
\end{equation}
where $c$ is to be determined - physically this means that we are looking for slowly evolving solutions on a time-scale of order
$1/\epsilon^2$ in a frame of reference traveling with speed $c=\mathcal{O}(1)$. 
All the derivatives with respect to $z$ and $t$ in the basic equations
are then rewritten in terms of derivatives with respect to $x$ and $\tau$ by using the
transformations
\begin{equation}
\frac{\partial}{\partial z}= \frac{\partial}{\partial x}, \qquad 
\frac{\partial}{\partial t}=-c \frac{\partial}{\partial x} +\epsilon^2 \frac{\partial}{\partial \tau}.
\label{chain}
\end{equation}
Substituting  (\ref{Sexp}) and (\ref{Vexp}) into (\ref{basic2}),  (\ref{basic4b}) and (\ref{basic4})
  gives at the lowest order
\begin{equation}
V_{0rr}+\frac{1}{r} V_{0r}=0,
\label{V1}
\end{equation}
with the boundary conditions
\begin{equation}
V_0=1 \quad {\hbox{on}} \quad r=D,  \qquad {\hbox{and}}  \qquad V_0=0 \quad {\hbox{on}} \quad r=1.
\label{V2}
\end{equation}
The solution of (\ref{V1}) and (\ref{V2}) is
\begin{equation}
V_0\equiv V_0(r)=\frac{\ln r}{\ln D}.
\label{VS0}
\end{equation}
At the  order $\epsilon^2$  we have 
\begin{equation}
V_{1rr}+\frac{1}{r} V_{1r}=0,
\label{V3}
\end{equation}
subject to the boundary conditions
\begin{equation}
V_1\big|_{r=1}=-\eta\, \frac{1}{\ln D}, \qquad {\hbox{and}} \qquad V_1\big|_{r=D}=0.
\label{V4}
\end{equation}
Note that the interfacial position depends on the slow time-scale, i.e. $\eta\equiv\eta(x,\tau)$.
The first of the conditions (\ref{V4}) follows from (\ref{basic4b})  after evaluating at the interfacial position given by (\ref{Sexp}) and
expanding to order $\epsilon^2$. The solution of (\ref{V3}) and (\ref{V4}) is
\begin{equation}
V_1=\frac{\eta \ln r}{\ln^2 D}-\frac{\eta}{\ln D}
\label{VS1}
\end{equation}
Proceeding to order $\epsilon^4$ we obtain the problem
\begin{equation}
V_{2rr}+\frac{1}{r} V_{2r}=-V_{1xx},
\label{V5}
\end{equation}
subject to the boundary conditions
\begin{equation}
V_2\big|_{r=1}=-V_{1r}\big|_{r=1}\, \eta-\frac{1}{2} V_{0rr}\big|_{r=1}\, \eta^2, 
\label{V6}
\end{equation}
and
\begin{equation}
V_2\big|_{r=D}=0.
\label{V7}
\end{equation}
Using the solutions (\ref{VS0}) and (\ref{VS1}) we can rewrite (\ref{V6}) as
\begin{equation}
V_2\big|_{r=1}=-\frac{\eta^2}{\ln^2 D}+\frac{\eta^2\, \ln D}{2}.
\label{V6b}
\end{equation} 
The solution of (\ref{V5}) subject to the boundary conditions (\ref{V7}) and (\ref{V6b}) is
\begin{equation}
V_2=-\frac{\eta_{xx}}{\ln^2 D} \left( \frac{r^2}{4}\ln r-\frac{r^2}{4}\right)
+\frac{r^2\,\eta_{xx}}{4\ln D} +A\,\ln r+B,
\label{VS2}
\end{equation}
where
$$
B=-\frac{\eta^2}{\ln^2 D} +\frac{\eta^2 }{2\ln D}-\frac{1}{4}\,
\frac{\eta_{xx}}{\ln^2 D}-\frac{\eta_{xx}}{ 4 \ln D}
$$
and 
$$
A=\frac{\eta_{xx}}{\ln^2 D} \left(\frac{D^2}{4}-\frac{1}{4}\, \frac{D^2}{\ln D}\right)
-\frac{D^2}{4\ln^2 D} \eta_{xx}-\frac{B}{ \ln D}.
$$

For the fluid dynamics we substitute the expansions \eqref{phiexp} into the Laplace equation \eqref{basic1}
and obtain the following solutions at the first three orders
\begin{align}
&\phi_0=\phi_0(x,\tau),\label{phi0}\\
&\phi_1=-\frac{r^2}{4}\,\phi_{0xx}+\theta_1(x,\tau),\label{phi1}\\
&\phi_2=\frac{r^4}{64}\,\phi_{0xxxx}-\frac{r^2}{4}\,\theta_{1xx}+\theta_2(x,\tau)\label{phi2},
\end{align}
where $\theta_1(x,\tau)$ and $\theta_2(x,\tau)$ are unknown functions. Note that all the terms that are singular
at $r=0$ have been dropped from the solutions above. 
The boundary conditions to be imposed at $r=S$ are the kinematic and
dynamic equations (\ref{eq:kin}) and (\ref{basic5}), respectively. 
The former gives at the order $\epsilon^4$
\begin{equation}
-\frac{1}{2} \,\phi_{0xx}=-c\,\eta_x,
\label{kin1}
\end{equation}
whereas the dynamic boundary condition (\ref{basic5}) gives at order $\epsilon^2$
\begin{equation}
-c\,\phi_{0x}+\eta \, \left(E_b \frac{\ln D-1}{ \ln^3 D}-1\right)=0.
\label{basic51}
\end{equation}
Eliminating $\phi_0$ between (\ref{kin1}) and (\ref{basic51}) by differentiating (\ref{basic51})
with respect to $x$ yields
\begin{equation}
-2c^2 \eta_x+ \big(E_b \frac{\ln d-1}{ \ln^3 d}-1\big) \eta_x=0,
\end{equation}
which implies that
\begin{equation}
c^2=\frac{1}{2} E_b \, \frac{\ln D-1}{ \ln^3 D}-\frac{1}{2}.
\label{defc}
\end{equation}
Equation (\ref{defc}) defines the velocity $c$ in terms of the basic variables. It requires $D>e$ and $E_b$ to be
sufficiently large so that the right hand side of (\ref{defc}) is positive. As expected, \eqref{defc} is identical
(after non-dimensionalization)
to the leading order long-wave dispersion relation given by \eqref{eq:disp}.

In order to find equations for the unknown $\eta(x,\tau)$, we proceed to higher order. The kinematic boundary 
condition (\ref{eq:kin}) gives at order $\epsilon^6$ 
\begin{equation}
\eta_{\tau}+\phi_{0x} \eta_x=\frac{1}{16} \phi_{0xxxx}-\frac{1}{2} \theta_{1xx}-\frac{1}{2} \phi_{0xx}\, \eta,
\label{kin2}
\end{equation}
whereas the dynamic boundary condition (\ref{basic5}) gives at the order $\epsilon^4$
$$
\frac{c}{4} \phi_{0xxx}-c\,\theta_{1x}+\phi_{0\tau}+\eta^2
-\eta_{xx}
-\frac{E_b \,\eta^2}{2\ln^4 D}\left(3-5\ln d+3\ln^2 d\right)
$$
\begin{equation}
-\frac{E_b\, \eta_{xx}}{4\ln^4 d}\left(1-D^2+2\ln D+2\ln^2 D\right)
+\frac{1}{2} \phi_{0x}^2=0.
\label{basic52}
\end{equation}
Next we eliminate $\theta_1$ between (\ref{kin2}) and (\ref{basic52})  by  first differentiating (\ref{basic52})
with respect to $x$. This yields
$$
2c\,\eta_{\tau}+2c\,\phi_{0x}\,\eta_x+\frac{c}{8}\,\phi_{0xxxx}+\phi_{0x\tau}
+(2-P_1)\, \eta\eta_x-(1+P_2)\,\eta_{xxx}
$$
\begin{equation}
+\phi_{0x}\, \phi_{0xx}+c\,\phi_{0xx}\,\eta=0,
\label{final0}
\end{equation}
where $P_1$ and $P_2$ are defined as
\begin{align}
P_1&=\left[\frac{3\ln^2D-5\ln D+3}{\ln^4D}\right]\,E_b,\\
P_2&=\left[\frac{2\ln^2D+2\ln D+1-D^2}{4\ln^4D}\right]\,E_b.
\end{align}
Relations (\ref{kin1}) and (\ref{basic51}) 
 imply that
\begin{equation}
\phi_{0x}=2c\,\eta,
\label{final1}
\end{equation}
and on substituting this into \eqref{final0} yields a single evolution equation for the jet shape $\eta(x,T)$:
\begin{equation}
\eta_{T} +\alpha \,\eta \eta_x+\beta\,\eta_{xxx}=0,
\label{final2}
\end{equation} 
where
\begin{equation}\label{constants}
\alpha=10c^2+2-P_1,\qquad\beta=\frac{c^2}{4}-P_2-1,
\end{equation}
and time has been rescaled according to $4c\,\partial_\tau\to\,\partial_T$ (this is possible since $c>0$).
The coefficients $\alpha$ and $\beta$  of the nonlinearity and dispersive terms, respectively, determine the
type of soliton solutions that are supported. The classical solitons studied in water waves (see Whitham \cite{Whitham},
for example) have $\alpha>0,\,\beta>0$ (e.g. $\alpha=1$, $\beta=6$) are functions of $\xi=x-sT$,
where $s$ is the wave speed; solutions with $\alpha<0$ are also found in waves problems with an internal interface
(see for example Dias \& Vanden-Broeck \cite{DiasVB} and references therein). 
Looking for such solutions in \eqref{final2} and integrating twice with respect to $\xi$ yields (we also use the
fact that $\eta(\xi)\to0$ as $|\xi|\to\infty$)
\begin{equation}
\beta \left(\frac{d\eta}{d\xi}\right)^2=s\eta^2-\frac{\alpha}{3}\eta^3.\label{eq:kdvInt}
\end{equation}
If the amplitude of the wave is $\eta_0$ (this can be positive or negative), it follows by evaluating \eqref{eq:kdvInt}
at the crest/trough that the speed $s$ is given by
\begin{equation}
s=\frac{1}{3}\alpha\eta_0,\label{eq:speed-s}
\end{equation}
and so \eqref{eq:kdvInt} takes the form
\begin{equation}
\beta \left(\frac{d\eta}{d\xi}\right)^2=\frac{\alpha}{3}\eta^2(\eta_0-\eta).\label{eq:kdvInt-1}
\end{equation}
It is shown below that $\beta>0$ is a necessary condition for solitary waves
to exist, and hence there are two possibilities: (i) $\alpha>0$ in which case $\eta<\eta_0$ with $\eta_0>0$, giving
waves of elevation, and (ii) $\alpha<0$ in which case $\eta>\eta_0$ with $\eta_0<0$, giving waves of depression.
From the expression \eqref{eq:speed-s} for the speed we conclude that both elevation and depression waves
have $s>0$ and hence their speed is supersonic relative to the linear speed $c$ (see \eqref{defc}).
The solutions of \eqref{final2} are the well-known solitary wave solutions
\begin{equation}
u(x,T)=\frac{3s}{\alpha}\,{\rm{sech}}^2\left[\frac{1}{2}\sqrt{\frac{s}{\beta}}\left(x-sT\right)\right],\label{soliton}
\end{equation}
and as discussed above we obtain right-moving elevation or depression solitons for $\alpha>0$ or $\alpha<0$, respectively.

\begin{figure*}
  \includegraphics[width=12cm, height=9cm]{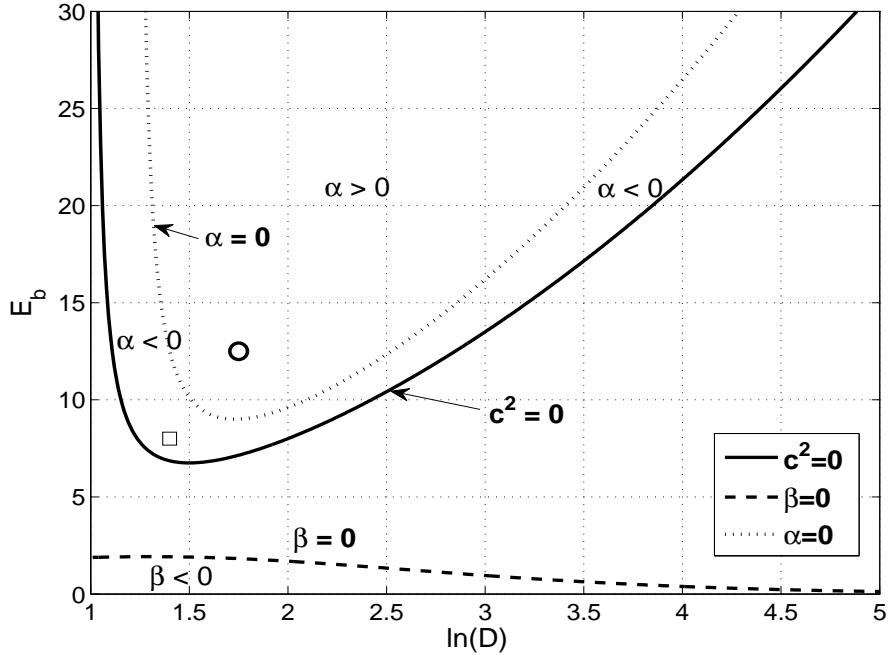}
  \caption{Different regions in the $E_b-D$ plane where admissible solitary waves can be found.
  The curves indicate where $c^2=0$ (solid), $\beta=0$ (dashed) and $\alpha=0$ (dotted);
  the corresponding values of the parameters are positive/negative above/below the curves. The open circle
  indicates the point $\ln D=1.75$, $E_b=12.5$ (i.e. $\alpha=1.649$, $\beta=6.5874$ in \eqref{final2}), and the
  square corresponds to $\ln D=1.4$, $E_b=8$ (i.e. $\alpha=-1.0841$, $\beta=3.5630$ in \eqref{final2}).
  The corresponding solitary waves are plotted in Figure \ref{fig:4}.}
  \label{fig:regions}
\end{figure*}

\begin{figure*}
  \includegraphics[width=12cm, height=9cm]{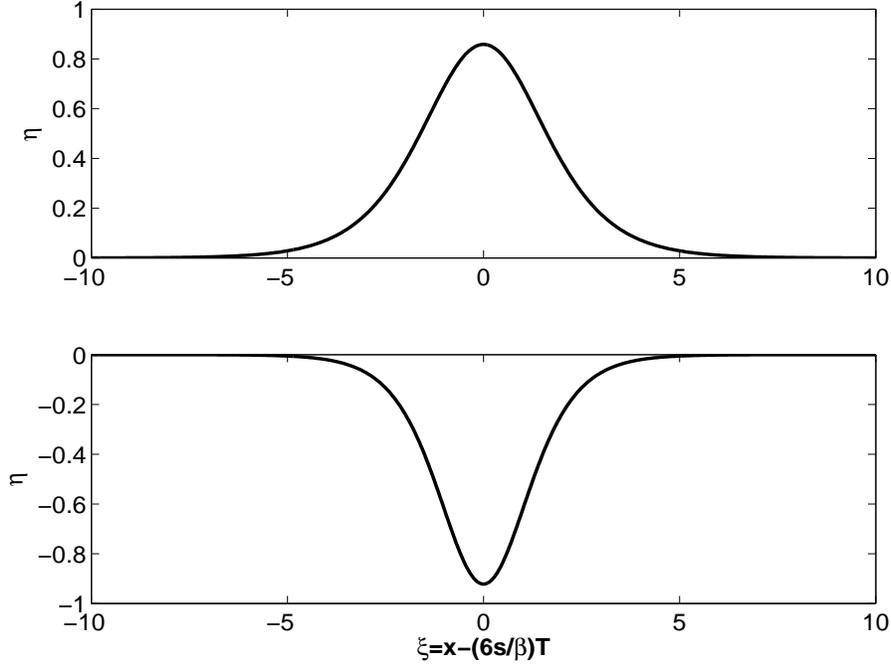}
  \caption{Elevation (top) and depression (bottom) of solitary waves corresponding to the open circle and square, respectively,
  on Figure \ref{fig:regions}; the parameter values are $\ln D=1.17$, $E_b=12.5$ and 
$\ln D=1.4$, $E_b=8$.}
  \label{fig:4}
\end{figure*}

It remains to determine whether equation \eqref{final2} with \eqref{constants} can support both elevation and depression solitary waves. 
The parameters $\alpha$ and $\beta$ are functions of the dimensionless outer electrode
radius $D$ and the electric field parameter $E_b$ (large $E_b$ implies a large imposed electric field). Equation \eqref{defc}
for the speed imposes the constraint $\ln D>1$, and so we consider such values alone.
In terms of $D$ and $E_b$ the expressions for $\alpha$ and $\beta$ are
\begin{eqnarray}
\alpha&=&\frac{E_b}{\ln^4 D}\left[ 2\ln^2 D-3\right]-3,\label{eq:alpha}\\
\beta&=&\frac{E_b}{8\ln^4 D}\left[2D^2-3\ln^2 D-5\ln D-2 \right]-\frac{9}{8}.\label{eq:beta}
\end{eqnarray}
To determine regions in $D-E_b$ space where $\alpha$ and $\beta$ are positive or negative, it is useful to plot
the curves $\alpha=0$, $\beta=0$. It can be seen from \eqref{eq:alpha} that if $\ln D\le \sqrt{\frac{3}{2}}$ then $\alpha<0$ for
all positive values of $E_b$ ($E_b<0$ is unphysical). The curves $\alpha=0, \beta=0$ are depicted in Figure \ref{fig:regions},
and $\alpha, \beta$ are positive or negative above and below the curves, respectively, as indicated on the figure. The additional constraint 
$c^2>0$ with the speed given by \eqref{defc}, restricts permissible values of $D$ and $E_b$. The curve where $c^2=0$
is also included in Figure \ref{fig:regions} (solid curve), and we have $c^2>0$ above it. 
We can conclude, therefore, that for the waves found here
we always have
$$
\beta>0,
$$
while there are regions in parameter space where $\alpha$ can be positive or negative producing
elevation or depression waves, respectively.
In particular, $\alpha<0$ in the region that lies between the curves $c^2=0$ and $\alpha=0$, as
indicated on Figure \ref{fig:regions}. Typical waves are plotted in Figure \ref{fig:4} for $\alpha>0$ (top panel) and $\alpha<0$
(bottom panel), respectively. These waves correspond to the pairs of parameter values $\ln D=1.17$, $E_b=12.5$ and 
$\ln D=1.4$, $E_b=8$; these points are depicted by an open circle and a square, respectively, on Figure \ref{fig:regions}. 

It is also worth noting that for values of $D$ and $E_b$ that give $\alpha=0$,
the quadratic nonlinearity in the equation vanishes and a different asymptotic analysis is required to produce
a higher order nonlinearity - the third derivative term cannot vanish as mentioned above - and so the system
that derives is a Kortweg de-Vries equation with a high order nonlinearity. This is not pursued further here
and is left for future work.

It is useful to transform the scaled Kortweg de-Vries equation \eqref{final2} back to original variables
in order to demonstrate the asymptotic balances of weak nonlinearity and weak dispersion.
Recalling the non-dimensionalizations \eqref{eq:scalings} and the asymptotic scalings $S=1+\epsilon^2\eta$
and time transformation $\partial_t=-c\partial_x+\epsilon^2\partial_\tau$ (see \eqref{Sexp} and \eqref{chain}),
the equation takes the following form in terms of dimensional variables
\begin{eqnarray}
S_t+c^* S_z+\frac{\alpha^*}{4c^*}\frac{(S-a)}{a} S_z+\frac{\beta^*}{c^*} a^2 S_{zzz}=0,\label{kdv-dim}
\end{eqnarray}
where $c^*=\left(\frac{\gamma}{\rho a}\right)^{1/2} c$ is the dimensional wave speed of linear
long waves (the dimensional version of $c$ given by \eqref{defc}), and $\alpha^*=\left(\frac{\gamma}{\rho a}\right)\alpha$ and 
$\beta^*=\left(\frac{\gamma}{\rho a}\right)\beta$ are dimensional parameters having units of velocity-squared and 
depending on the electric field and the geometric ratio $a/d$ - see \eqref{eq:alpha}-\eqref{eq:beta}. In the long wave
limit and for weakly nonlinear perturbations, equation \eqref{kdv-dim} shows clearly the balances between
nonlinearity and weak dispersion (after moving to a frame of reference of speed $c^*$ and introduction of
a slow timescale as described in detail for the dimensionless equations in Section \ref{sec:weakly}).

\section{Conclusions}
\label{conclusions}
We have considered the weakly nonlinear evolution of long wave axisymmetric disturbances on a cylindrical liquid
jet under the influence of a radially imposed electric field. We find that at sufficiently large imposed electric fields measured
by the electric Bond number $E_b$ (see the definition just after equation \eqref{invdisp0}), and above a critical
outer electrode radius $D=d/a>e\approx 2.7183$, long waves are dispersive and an asymptotic analysis analogous to
that used to derive the Korweg de-Vries equation for water waves (see Whitham \cite{Whitham}) is applicable
and leads to the KdV equation \eqref{final2}. The coefficients $\alpha$ and $\beta$ of the nonlinear and dispersive terms, respectively,
depend on the two parameters $E_b$ and $D$, and admissible values are additionally constrained by the condition $c^2>0$ where
the latter is given by the formula \eqref{defc}. It is found that for the theory to hold we must have $\beta>0$, whereas $\alpha$ can
be positive or negative as indicated in the phase diagram in Figure \ref{fig:regions}. If $\alpha>0$ we obtain right moving
solitary waves of elevation while for $\alpha<0$ depression solitons emerge - see Figure \ref{fig:4} for representative solutions.
We note that similar waves were found in the case of ferrofluids (see \ref{introduction}) but interestingly the mathematical
problems are quite different; in the present problem a Laplace equation for the electric field must be solved in the annulus
in order to find the appropriate term in the Bernoulli equation, whereas for the ferrofluid problem there is a decoupling
and the Bernoulli equation is simply modified by a term inversely proportional to the local jet radius. We also note that 
the solutions constructed here are likely to be susceptible to shorter wave disturbances that would be modulated by
the soliton envelope - such calculations are beyond the scope of the present work and would most likely require
time-dependent direct numerical simulations. This is left for future work.


\begin{acknowledgments}
The work of D.T.P. and J.-M.V.B. was partly supported by the Engineering and Physical Sciences Research Council of Great Britain by grant
numbers EP/K041134/1 and EP/J019569/1.
\end{acknowledgments}

\bibliographystyle{plain}
\bibliography{ElecJets}

\end{document}